\documentclass[aps,prl,final,twocolumn,superscriptaddress]{revtex4}

\usepackage{amssymb}

\usepackage[pdftex]{graphicx}
\usepackage{footmisc}
\graphicspath{{figures/}}
\usepackage{color}

\newcommand{\zrte}{ZrTe$_3$}
\newcommand{\zrtese}{ZrTe$_{2.96}$Se$_{0.04}$}

\newcommand{\bbar}{$\overline{\mathrm{B}}$}
\newcommand{\dbar}{$\overline{\mathrm{D}}$}
\newcommand{\ybar}{$\overline{\mathrm{Y}}$}
\newcommand{\gammabar}{$\overline{\mathrm{\Gamma}}$}

\newcommand{\invA}{\mathrm{\AA}^{-1}}
\renewcommand{\deg}{^\circ}

\begin{document}

\title{Disorder quenching of the Charge Density Wave in ZrTe$_3$}

\author{Moritz Hoesch}
\affiliation{Diamond Light Source, Harwell Campus, Didcot OX11 0DE, United Kingdom}
\affiliation{Hiroshima Synchrotron Radiation Center, Hiroshima University, 2-313 Kagamiyama, Higashi-Hiroshima 739-0046, Japan}

\author{Liam Gannon}
\affiliation{Diamond Light Source, Harwell Campus, Didcot OX11 0DE, United Kingdom}
\affiliation{Clarendon Laboratory, University of Oxford Physics Department, Parks Road, Oxford, OX1 3PU, United Kingdom}

\author{Kenya Shimada}
\affiliation{Hiroshima Synchrotron Radiation Center, Hiroshima University, 2-313 Kagamiyama, Higashi-Hiroshima 739-0046, Japan}

\author{Benjamin Parrett}
\affiliation{Diamond Light Source, Harwell Campus, Didcot OX11 0DE, United Kingdom}
\affiliation{London Centre for Nanotechnology and Department of Physics and Astronomy,
University College London, Gower Street, London WC1 E6BT, UK}

\author{Matthew D. Watson}
\affiliation{Diamond Light Source, Harwell Campus, Didcot OX11 0DE, United Kingdom}

\author{Timur K. Kim}
\affiliation{Diamond Light Source, Harwell Campus, Didcot OX11 0DE, United Kingdom}

\author{Xiangde Zhu}
\altaffiliation{Present and permanent address: High Magnetic Field Laboratory, Chinese Academy of Sciences - Hefei 230031, PRC}
\affiliation{Condensed Matter Physics and Materials Science Department, Brookhaven National Laboratory Upton, NY 11973, USA}

\author{Cedomir Petrovic}
\affiliation{Condensed Matter Physics and Materials Science Department, Brookhaven National Laboratory Upton, NY 11973, USA}

\date{\today}

\begin{abstract}
The charge density wave (CDW) in \zrte\ is quenched in samples with small amount  of Te isoelectronically substituted by Se. Using angle-resolved photoemission spectroscopy we observe subtle changes in the electronic band dispersions and Fermi surfaces on Se substitution. The scattering rates are substantially increased, in particular for the large three-dimensional Fermi surface sheet.  The quasi-one-dimensional band is unaffected by the substitution and still shows a gap at low temperature, which starts to open from room temperature. The detailed temperature dependence reveals that the long-range order is absent in the electronic states as in the periodic lattice distortion. The competition between superconductivity and CDW is thus linked to the suppression of long-range order of the CDW.
\end{abstract}

\maketitle

The charge density wave (CDW) is a much studied self-organisation of metallic electrons in a crystalline solid~\cite{grunerbook, monceau12}. Besides  electronic signatures it manifests itself as periodic lattice distortion (PLD) with a periodicity given by the CDW modulation. In the last decade the competition between CDWs and superconductivity (SC) has gained renewed attention, due to experiments that include observations of incommensurate PLDs in the copper oxide family of superconductors~\cite{grininghellli12,chang12,tacon2013}. In a very simple picture a competition between CDW and SC may arise due to the removal of spectral weight from the Density of States at the Fermi level (DOS at $E_F$) if an energy gap is formed by the CDW transition.  Quenching the CDW restores DOS at $E_F$, which is required for SC. If the quenching arises due to static disorder this will affect the superconducting properties as well, but these have been shown theoretically to be more robust, especially in the presence of electron corralations~\cite{anderson1959,tang2016}.  In the extreme case of Josephson-linked networks of 1-dimensional chain segments, superconductivity is even found to be enhanced by disorder~\cite{petrovic2016}.

Besides the complex oxides cited above, competition between CDW and SC exists in rather simple binary materials. For example quenching of a CDW state under pressure and concomitant emergence of superconductivity has been found widely in stoichiometric transition-metal trichalcogenides~\cite{nunez1993,yomo05} and rare-earth trichalcogenides~\cite{sachetti2009,hamlin2009,zocco2015} as well as layered transition-metal dichalcogenides~\cite{li2016}. The specific arrangement of electronic bands that form the CDW as well as the PLD vary widely between these materials thus establishing the competition as a fundamental principle of solid state physics. The closest proximity of two transition temperatures at ambient pressure is found in 2H-NbSe$_2$ ($T_c =7.2$~K~\cite{revolinsky65}, $T_{CDW}=33$~K~\cite{wilson2013}). 

\begin{figure}
\centerline{\includegraphics[width = .46\textwidth]{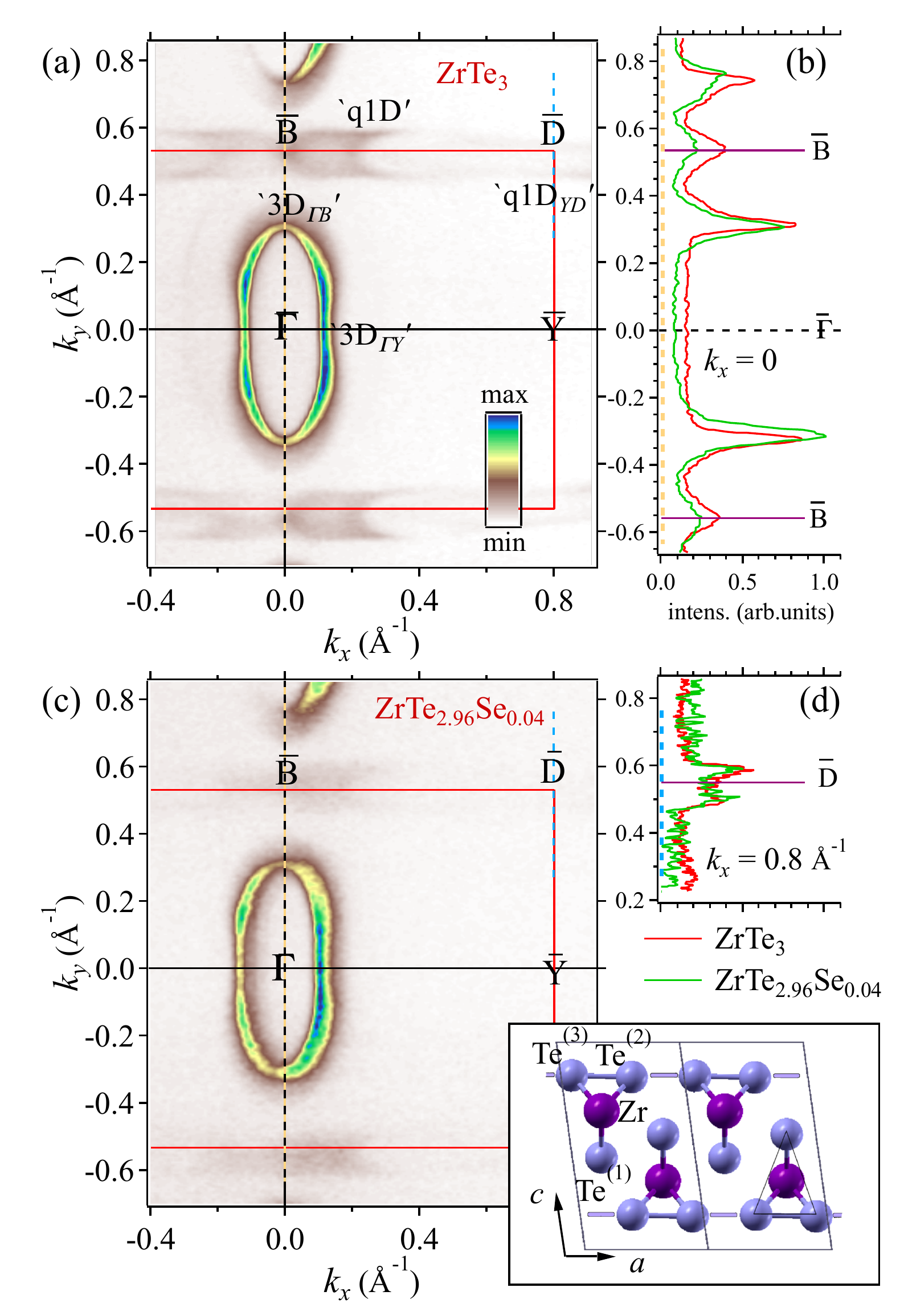}}
\caption{FS maps of \zrte\ (a) and \zrtese\ (c), acquired by ARPES at a temperature  $T = 7$~K. The inset shows two unit cells of the crystal lattice of \zrte\ projected onto the $(a,c)$-planes. The nearly equidistant Te chains adjacent to the van der Waals gap are highlighted by grey bars. Panels (b) and  (d) show momentum distribution curves extracted from the maps along straight lines along $k_y$ at two different $k_x$.}
\label{Fig1}
\end{figure}

The uniaxial material \zrte\ shows (filamentary) SC below about $T_c \simeq 2$~K at ambient pressure~\cite{takahashi83, tsuchiya2017}. The CDW is seen as a resistivity anomaly around $T_{CDW} = 63$~K~\cite{takahashi83} with an incommensurate PLD modulation $\vec{q}_{CDW}\simeq (0.07,0,0.333)$~\cite{eaglesham84}. The resistivity continues to drop in a metallic fashion below  $T_{CDW}$. The two main FS sheets are a hole-like sheet in the $(a,b)$-plane (3D) and a pair of electron-like and very flat sheets from quasi-one dimensional (q1D) bands~\cite{felser98,stowe98,yokoya05}. The latter are formed by Te~5$p$-electrons of the Te$^{(2)}$ and Te$^{(3)}$ sites being adjacent to the van-der-Waals gap of the crystal structure (inset in Fig.~\ref{Fig1}), while the 3D sheet has a dominant Zr~4$d$ character. Bulk superconductivity emerges in \zrte\ when the CDW is quenched, {\em e.g.} at hydrostatic pressures above $P_c  =5$~GPa~\cite{yomo05} or in disordered samples grown at high temperatures~\cite{zhu13}. For the quenching of the CDW under pressure two potential mechanisms have been discussed: disorder~\cite{gleason15,kwang-hua2012}, or a re-arrangement of band fillings in the multi-sheet Fermi surface (FS) leading to the loss of the CDW stability~\cite{hoesch2016}. In either scenario it is reasonable to assume that the SC  involves electrons in the quasi-one-dimensional (q1D) sheets of the FS that have been identified as driving the CDW~\cite{felser98, stowe98,yokoya05} and which show strong electron-phonon coupling to low-energy vibrational modes~\cite{hoesch09ixs,hu2015}. This electron-lattice interaction probably contributes to stabilising the long wavelength (small $q$) CDW. 

In this letter we access the regime of low disorder by light chemical substitution of Te with Se, ZrTe$_{\mathrm{3-x}}$Se$_\mathrm{x}$. A suppression of the CDW both in transition temperature and magnitude of the resistivity anomaly was reported~\cite{zhu16}. Above $x=0.02$ the CDW anomaly is removed and bulk superconductivity with $T_c$ up to 4.5~K emerges. With the  isoelectronic substitution only a slight rearrangement of carriers between FS sheets is expected.  From the data presented below we clarify these subtle changes and find that even the slight disorder in a sample of $x=0.04$, exchanging less than 1.5\% of Te, leads to strong and sheet-dependent decreases in electron scattering length and a shift of the van-Hove-singularity (vHS) in the q1D band to occupied states. The signatures of charge density wave gaps  in \zrte\ and effects of fluctuating CDW in the substituted samples  are observed.

Single-crystals of \zrte\ and \zrtese\ have been grown by chemical vapour transport and characterised as described before~\cite{zhu16}. For ARPES measurements, the samples were cleaved in ultrahigh vacuum at temperature $T = 7$~K and a vacuum of 10$^{-10}$~mbar, exposing (001) surfaces as the samples cleave in the van-der-Waals gap between layers formed by Te$^{\mathrm{(2)}}$-Te$^{\mathrm{(3)}}$ atoms (inset in Fig.~\ref{Fig1}). ARPES data were acquired at beamline I05-ARPES at Diamond Light Source~\cite{hoesch17} at $T$ between 7~K and 290~K. The photon energy was set to $h\nu = 34$~eV and the energy resolution was chosen as $\delta E =7$~meV with an angular resolution of less than $0.2\deg$. The polarisation vector was kept in the $(b,c)$-plane of the crystal lattice ($p$-polarised).

Fig.~\ref{Fig1} shows two FS maps for samples of pure \zrte\ and \zrtese. The labelling of high symmetry points  acknowledges the effectively 2-dimensional electronic state  observed in ARPES, which is assigned to a surface relaxation~\cite{hoesch09}. This relaxation also leads to a splitting of observed bands, thus complicating the analysis. We selected $h\nu=34$~eV, at which one component of the split bands dominates the spectra. At different $h\nu$ the same features are seen~\cite{hoesch09}, but strong contributions from both components would complicate the peak fitting analysis.  Band dispersions are observed by step-wise rotating the sample around $a$ for $k_x$ and by the angle-parallel detection of the hemispherical electron analyser for $k_y$.

The key features of the FS are found to be nearly identical between the two samples, namely the almond shaped central feature, labelled 3D. The manifold of q1D sheets is barely visible, close to \dbar\ due to suppression of spectral weight at the chemical potential ($E_F$) at low $T$ in both samples. Close to \bbar\ at $k_x=0.2$~\AA\ two closing contours of this sheet are nevertheless  clearly discernible. In \zrtese\ the visibility of this manifold is reduced but key dispersion features are seen near identical.

The data are analysed by slicing into momentum distribution curves (MDCs) at constant $E$ along a straight line in momentum space such as shown in Fig.~\ref{Fig1}(b, d) for $E=E_F$. Peaks in these curves correspond to band dispersion positions, their width corresponds to the inverse scattering lengths~\cite{baumberger2004}. Numerical analyses are performed by recording the position, width and intensity from successive MDCs with increasing $E$ up to $E=E_F$. When two equivalent bands are included in the fit the width is forced to be identical for both, while positions and intensities are allowed to converge freely. 

\begin{figure*}[th]
\centerline{\includegraphics[width = .92\textwidth]{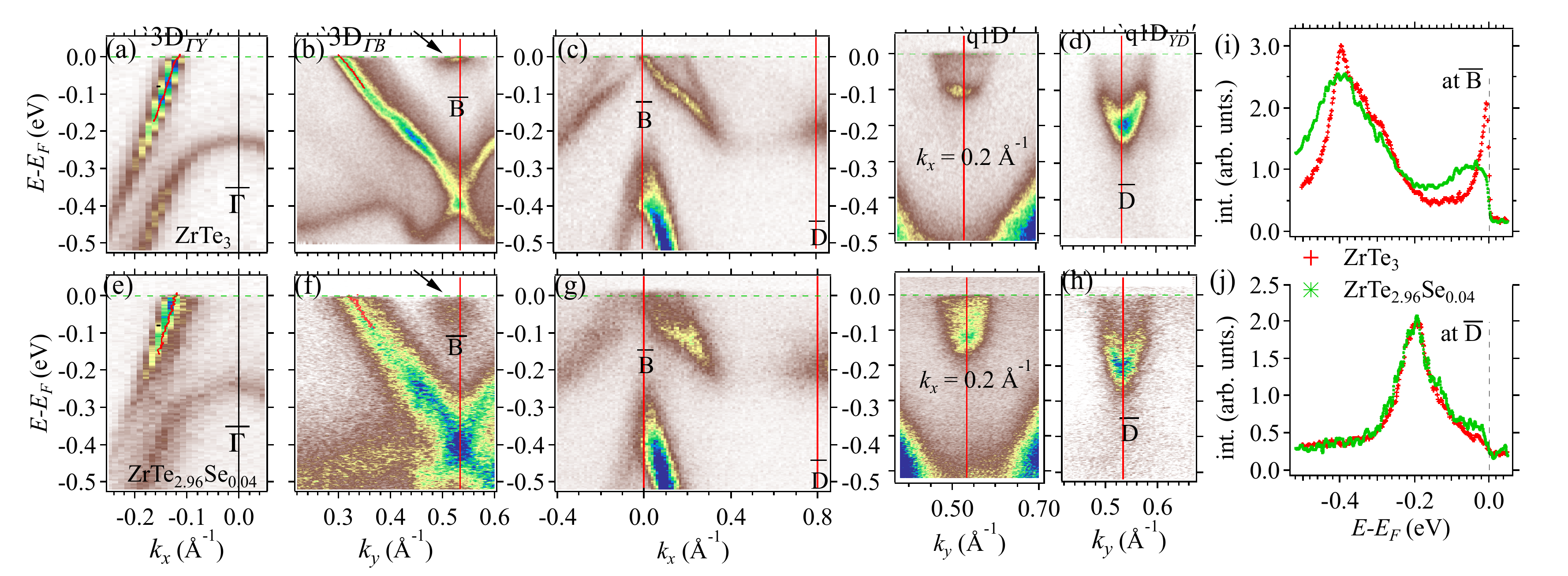}}
\caption{ARPES dispersion cuts for \zrte\ (a, b, c, d) and \zrtese\ (e, f, g, h) at $T=7$~K. Red dots in panels (a, b, e, f) show the peak positions of momentum distribution curves (MDCs). Note that the stepped look of the data in (a) and (e) is a graphical representation artefact that arises despite of slight over-sampling along the angle-scanned acquisition direction $k_x$. Energy distribution curves (EDCs) at  high symmetry points are shown in (i) for \bbar\ and in (j) for \dbar\ for both samples.}
\label{Fig2}
\end{figure*}

We first keep our focus on the FS shapes and anisotropies.
Most remarkably, the Fermi wave vectors of the q1D band close to \dbar\  are found to be identical within the error bars between both samples [Fig.~\ref{Fig1}(d)]. Similarly, the $k_x$ Fermi wave vectors along \gammabar-\ybar\ are nearly unchanged, though close analysis reveals an increase by 2\%. The $k_y$ Fermi wave vectors of the same sheet along \gammabar-\bbar\ on the other hand are reduced by about 4\%, which is easily visible in Fig.~\ref{Fig1}(b).The  Fermi wave vectors $k_F$ are summarise in Tab.~\ref{tab1}. The anisotropy of the 3D sheet is thus slightly reduced and the FS volume is decreased in \zrtese\  when compared to  \zrte. From this reduction of FS volume we can estimate a reduction of hole-type carrier numbers in the 3D band of about 2\%.

Fig.~\ref{Fig2} shows the ARPES dispersion cuts (energy-momentum maps) for both samples. The 3D bands [panels (a, b, e, f)] show near identical dispersions, apart from a momentum shift corresponding to the change of $k_F$ noted above. The biggest change is seen at precisely the \bbar\ point at $|E-E_F|<0.1$~eV [Fig.~\ref{Fig2}(b and f), marked by an arrow]. This peak derives from a vHS within $|E_{vH}-E_F| \simeq 0.01$~eV in \zrte~\cite{yokoya05}, which is shifted to occupied states in \zrtese. By tracing the small dispersion of the q1D band along $k_x$ [Fig.~\ref{Fig2}(g)] we find $E_{vH}-E_F = -0.03\pm0.005$~eV at \bbar\ in \zrtese. The spectra at \bbar\ are also reproduced in Fig.~\ref{Fig2}(i) showing the sharp peak of the vHS much broader in \zrtese. 

At \dbar\ the intensity distributions from either sample are remarkably similar with near identical spectra shown in Fig.~\ref{Fig2}(j), which are integrated over a small  range $0.4 < k_y < 0.67$~\AA$^{-1}$. Data from both samples show a peak at $E-E_F=-0.19$~eV with a characteristic reduction of intensity towards $E_F$. Close inspection shows a  higher intensity at $E=E_F$ for \zrtese\ when compared to \zrte. The $T$ dependence of this intensity will be analysed below to gain further insight on its relation to CDW formation. 

\begin{figure}
\centerline{\includegraphics[width = .5\textwidth]{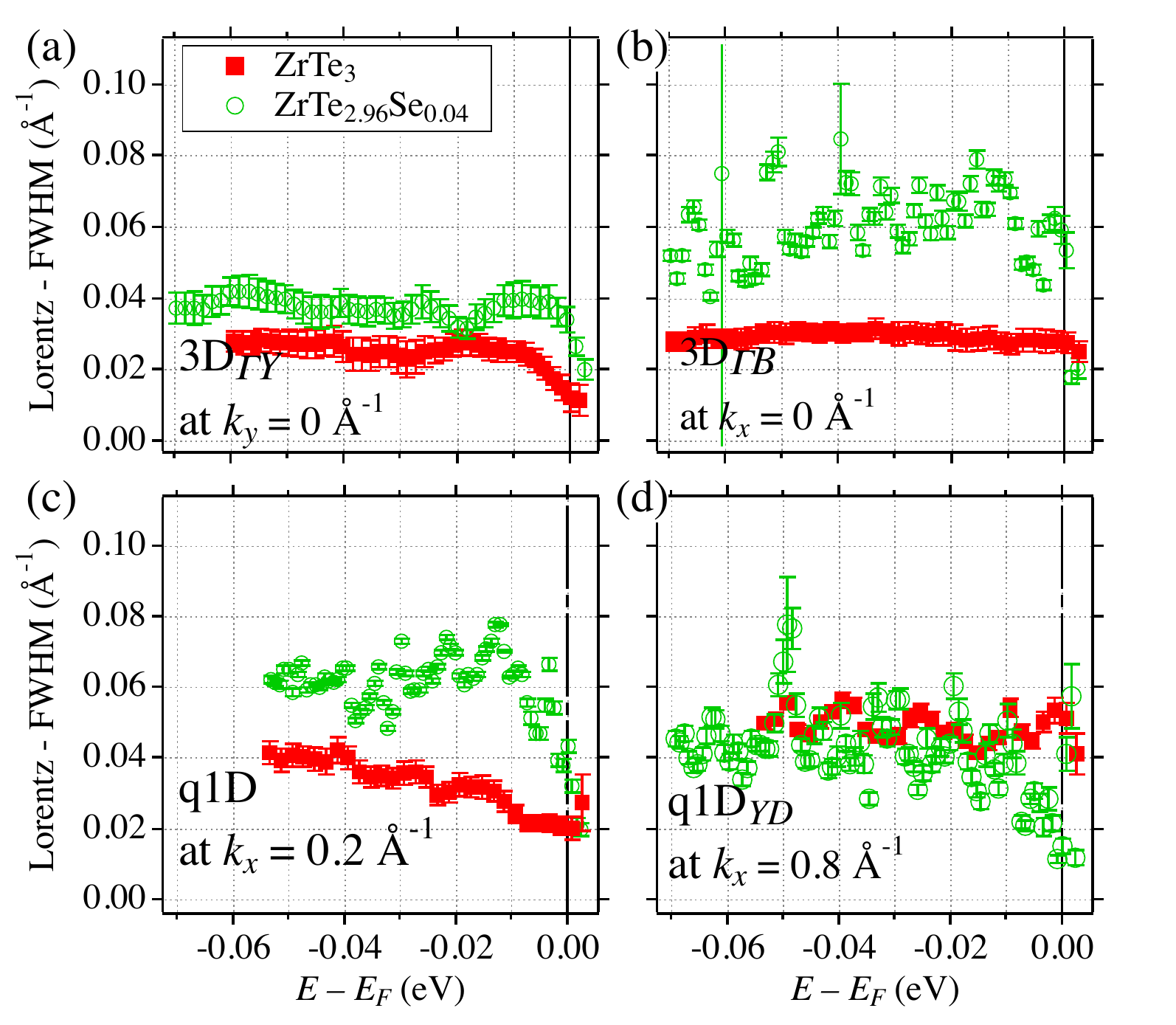}}
\caption{Extracted peak widths from fits to MDCs as a function of $E$ for both samples. The corresponding peak positions are shown in Fig.~\ref{Fig2}. Panel (a) shows the 3D band along $\Gamma-Y$, at $k_y=0$. Panel (b) shows the 3D band along $\Gamma-B$, at $k_x=0$. Panel (c) shows the q1D band along $k_y$ for $k_x = 0.2$~\AA$^{-1}$ and (d) near \dbar.}
\label{Fig3}
\end{figure}

The results of MDC peak widths as a function of $E$ are shown in Fig.~\ref{Fig3}. These widths can be considered as being composed of two main contributions: a constant width due to impurity scattering and an energy-dependent width due to electron-phonon coupling (EPC), which is zero at $E=E_F$~\cite{jiang2011}. The width due to EPC increases and saturates with decreasing $E$, in our data around $E-E_F=-0.01$~eV, which matches the range of phonons with strong EPC in \zrte~\cite{hoesch09ixs,hu2015}. For example a pronounced width increase by 0.012~\AA$^{-1}$ from 0.013~\AA$^{-1}$ (0.025~\AA$^{-1}$) is observed at position  `3D$_{\Gamma Y}$' for \zrte (\zrtese) [Fig.~\ref{Fig3}(a)]. Thus the effect of phonon scattering is similar, while the defect scattering length is decreased by a factor two in \zrtese\ when compared to \zrte. Along $k_y$, at  `3D$_{\Gamma B}$' as well as `q1D' (at $k_x=0.2$~\AA) the data for \zrtese\ show significantly higher widths throughout when compared to \zrte\ [Fig.~\ref{Fig3}(b,c)]. Near the \dbar\ point `q1D$_{YD} $' we find basically identical data for both samples, as seen already from the match of MDCs [Fig.~\ref{Fig1}(d)]. The thus extrapolated widths at $E_F$ due to defect scattering are summarised in Tab.~\ref{tab1}. When multiplied with the Fermi velocity $v_F$ they can give an estimate of the scattering rate $1/\tau$ due to disorder. The scattering rate is enhanced upon Se substitution except for position `q1D$_{YD}$' of the strongly nested CDW band~\footnote{We note that also in \zrte\ no linewidth broadening due to CDW formation is obseved near \dbar, as seen by the smooth evolution across $T_{CDW}$ in supplemental Figures S2 and S3~\cite{supplemental}}.

\begin{table}
	\centering
	\caption{Extracted values for the Fermi wave vector $k_F$, the peak width $\delta k$, the Fermi velocity $v_F$ and the thus estimated scattering rates for both samples. The upper part shows the data for \zrte, the lower part reports data for \zrtese.\vspace{3mm}}
	\label{tab1}
	\begin{tabular}{l||c|c|c|c}
	         \hline
	         \  band            &$2 k_F$ ($\invA$)& $\ \delta k$ ($\invA$)\ & $v_F$ (eV$\cdot\mathrm{\AA}$)& $1/\tau$ (eV) \\ 
		 \hline
		 \multicolumn{5}{l}{\hspace{1.2mm} \zrte} \\
		 \hline
		\  `3D$_{\Gamma Y}$'            &       0.633(3)           &             0.013(3)           &    4.0(2)  & 0.05(1) \\ 
		\  `3D$_{\Gamma B}$'           &       0.232(2)           &          0.026(4)              &   1.8(2)   & 0.05(1) \\
		\  `q1D'                                    &        0.12(2)           &           0.02(1)              &    4.8(9)    &  0.10(5) \\
		\  `q1D$_{YD}$'                      &        0.086(1)          &           0.05(2)              &    8.9(5)    & 0.44(20)  \\
		\hline
		 \multicolumn{5}{l}{\hspace{1.2mm} \zrtese} \\
		\hline
		\  `3D$_{\Gamma Y}$'            &      0.622(5)            &            0.025(5)           &     4.5(2)   & 0.11(2) \\ 
		\  `3D$_{\Gamma B}$'            &      0.244(3)            &            0.050(9)           &     2.1(2)   & 0.10(2) \\
		\  `q1D'                                    &       0.13(2)            &             0.04(1)           &   6.4(9)       & 0.26(7)\\
		\  `q1D$_{YD}$'                      &       0.085(1)           &             0.03(2)          &     9.2(5)     & 0.28(18) \\
		\hline
	\end{tabular}
\end{table}

\begin{figure}
\centerline{\includegraphics[width = .33\textwidth]{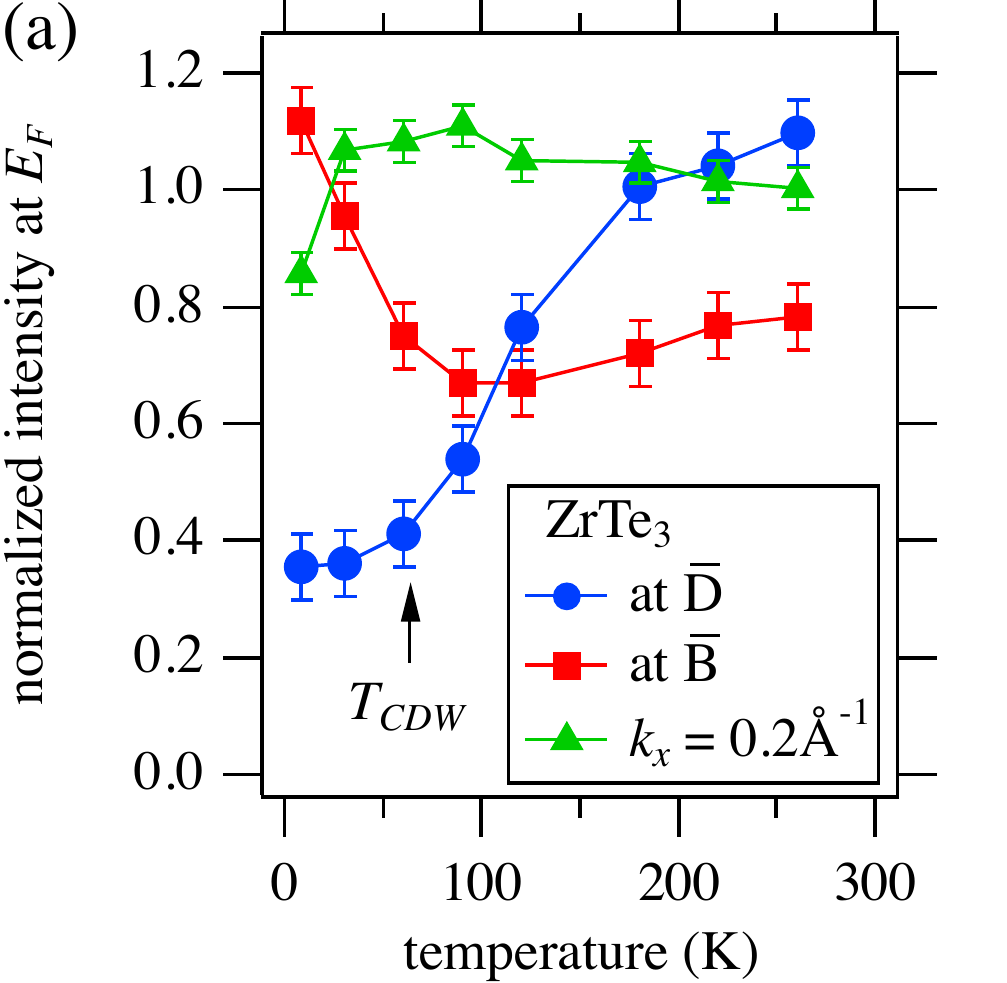}}
\centerline{\includegraphics[width = .33\textwidth]{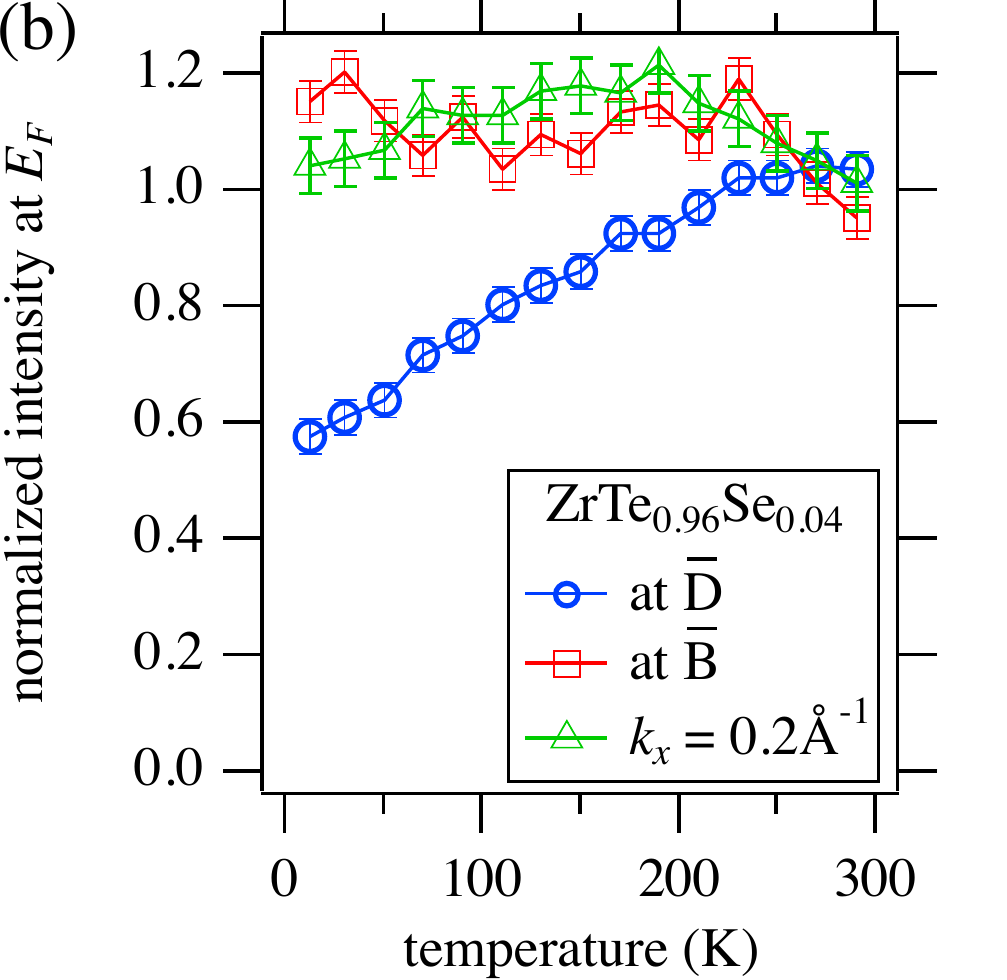}}
\caption{Temperature dependence of the photoemission intensity at $E_F$ from \zrte\ (a)  and \zrtese\ (b). The  intensities are summed up over a small range of $0.4 < k_y<0.67\ \invA$ at three different positions of $k_x$ as indicated.}
\label{Fig4}
\end{figure}

Analysing the intensity at $E_F$ as a function of $T$ (Fig.~\ref{Fig4}) we find near constant values for the 3D band (not shown) as well as for q1D near $k_x =0.2$~\AA$^{-1}$ (apart from a spurious lower intensity data point in \zrtese\ at the lowest $T$). This is regular behaviour of the Fermi-Dirac distribution function without any modification of the bands or spectral weight. The previously reported changes of intensity near \dbar\ and at \bbar\ are again found in our data form \zrte~\cite{yokoya05}. However, \zrtese\ does not show the increase at low $T$ at \bbar\ thus confirming the relation of this effect to the vHS. Near \dbar\ the  drop of intensity sets in at room temperature (RT) far above the structural transition $T_{CDW}$ in both samples.

From the data presented above a quenching of the CDW due to a rearrangement of bands in \zrtese\ is highly unlikely as the q1D band shows only minor changes. A slight increase of carrier numbers in this band is manifest from the lowering of binding energies near \bbar\, which was also observed theoretically at elevated pressures~\cite{starkovicz07,hoesch2016}. Considering the strong increase in scattering rates throughout we can conclude that the CDW in \zrtese\ is quenched by effects of static disorder.

A fluctuating CDW is still inferred from the depletion of spectral weight near \dbar. In \zrtese\ this follows a near linear trend with $T$ [Fig.~\ref{Fig4}(b)]. In \zrte\ the depletion is stronger, removing 2/3 of spectral weight at $E_F$ from the RT value, and the long-range ordering is manifest as a kink in the curve at $T_{CDW}$ [Fig.~\ref{Fig4}(a)]. This region of the Brillouin zone was shown to couple strongly to phonons with large and $T$-dependent width in Raman spectra~\cite{hu2015}. As the Fermi surface topography is basically unchanged here, we may speculate that a contribution to the CDW quenching comes from the chemical disorder disturbing the phonons that help to stabilise the PLD.  No long-range order is observed down to $T=20$~K in x-ray diffraction~\cite{supplemental}.

In conclusion, we find that the quenching of long-range CDW order with light isoelectronic substitution in \zrtese\ is driven by a decrease in the static scattering length. Our measurements thus confirm  the structural disorder as the primary tuning parameter which dictates the phase diagram of ZrTe$_{\mathrm{3-x}}$Se$_\mathrm{x}$. The effects of disorder in the electronic spectral function are found to be sheet-dependent, detected as a significant increase in the linewidths of the 3D band and the q1D bands near \bbar, while no measurable additional broadening was found near \dbar. In addition, more subtle band shifts are observed, including a shift of the vHS at \bbar\ away from the Fermi level to occupied states. A continuous partial loss of DOS around \dbar\ as a function of $T$ is interpreted as a signature of a fluctuating CDW as in previous work~\cite{yokoya05}. This is expected to weaken the SC and the shift of the vHS will also reduce the DOS at $E_F$ when compared to \zrte. The reduction of DOS from the region around \dbar\ is less strong in \zrtese\ than in \zrte\ and lacks the kink signature of long-range order that is observed in \zrte. Thus SC can emerge despite CDW fluctuations.

\section{Acknowledgments}
We are grateful to F. Baumberger for use of his ARPES data analysis software and to Diamond Light Source, where access to beamline I05 (NT11039, SI13797, and NT17065) and beamline I19 (MT8776) contributed to the results presented here. We acknowledge technical assistance for the ARPES experiments by Z.~K. Liu and L.~C. Rhodes and help with x-ray diffraction experiments was provided by S. Barnes, H. Novell, F. Fabrizi and D. Allen. Work at Brookhaven National Laboratory was supported by the U.S. Department of Energy, Office of Science, Office of Basic Energy Sciences, under Contract No. DE-SC0012704.

\newpage
\onecolumngrid

\section{Supplemental Material}

\subsection{Diffuse x-ray scattering}

The absence of a Charge Density Wave (CDW) signature was previously observed in electrical resistivity measurements~\cite{zhu16}. Here, we report data of X-ray diffraction experiments at a temperature of $T=20$~K that are sensitive to the periodic lattice distortion (PLD), which shows up as incommensurate superlattice reflextions at positions $\vec{Q}=\vec{G}\pm\vec{q}$ with $\vec{q}\sim(0.07,\ 0 ,\ 0.3333)$ in reciprocal lattice units (r.l.u.), where $\vec{G}$ is a main lattice Bragg spot position. The superlattice scattering intensity is weak compared to the main lattice scattering and  the CCD detector was deliberately over-exposed for these measurements, leading to large, slightly blurred intensity around the main lattice Bragg spots due to sample mosaic imperfections and residual thermal diffuse scattering (TDS). The experiments were performed at beamline I19 of Diamond Light Source.

\begin{figure*}[h!]
\centerline{\hspace{4mm}\includegraphics[height=4.5cm]{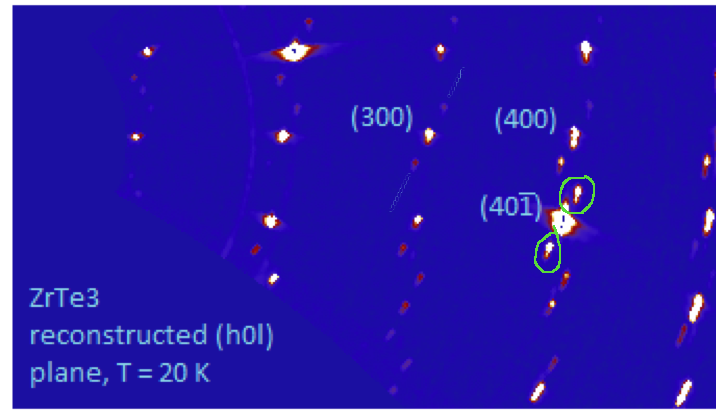}\hfill\includegraphics[height=4.5cm]{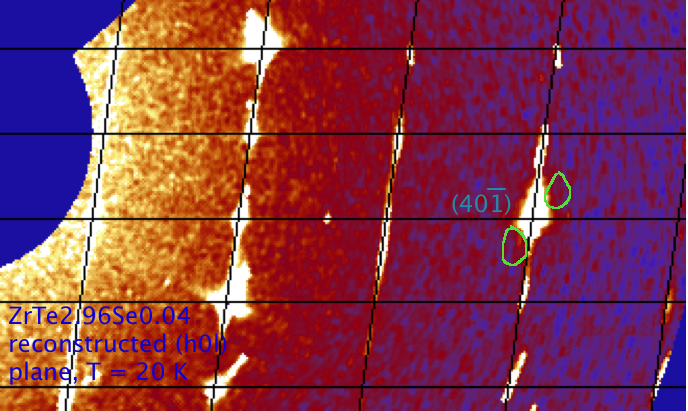}\hspace{4mm}}
\caption{Diffraction plane $(h0l)$ spanning approximately $0.2<h<5$ and  $-3 < l <1$ for pure \zrte\ (left) and \zrtese\ (right). Both data sets are constructed from a large number of sequential detector exposures while rotating the sample. The data are acquired at a temperature of $T = 20$~K. Key features for orientation on the maps are highlighted and green circles mark the positions of superstructure reflections due to the periodic lattice distortion, which are strong in \zrte\ and not observed in \zrtese.}
\label{FigS1}
\end{figure*}

Figure~S\ref{FigS1} shows the x-ray scattering intensity in the plane $\vec{Q}=(h,0,l)$ and spanned by $h$ (increasing to the right) and $l$ (increasing upward). The image has been reconstructed from a few dozen of detector exposures at varying sample angle. Data in panel (a) are for pure \zrte\ and show clear superstructure spots close to several main lattice positions, in particular close to $\vec{G} = (4,\ 0,\ \overline{1})$, where they are marked by green circles. Similar data have been reported in Ref.~[\onlinecite{hoesch2016}], where the PLD was reliably observed up to the quenching pressure of 5~GPa and absent above this pressure. Data from \zrtese\ shown in panel (b) show a smooth nearly constant level of background intensity around these superstructure positions. Therefore no long-range ordered PLD was observed in \zrtese\ in agreement with the transport results~\cite{zhu16}.

\subsection{Temperature-dependent ARPES}

\begin{figure*}[h!]
\centerline{\includegraphics[height=11.5cm]{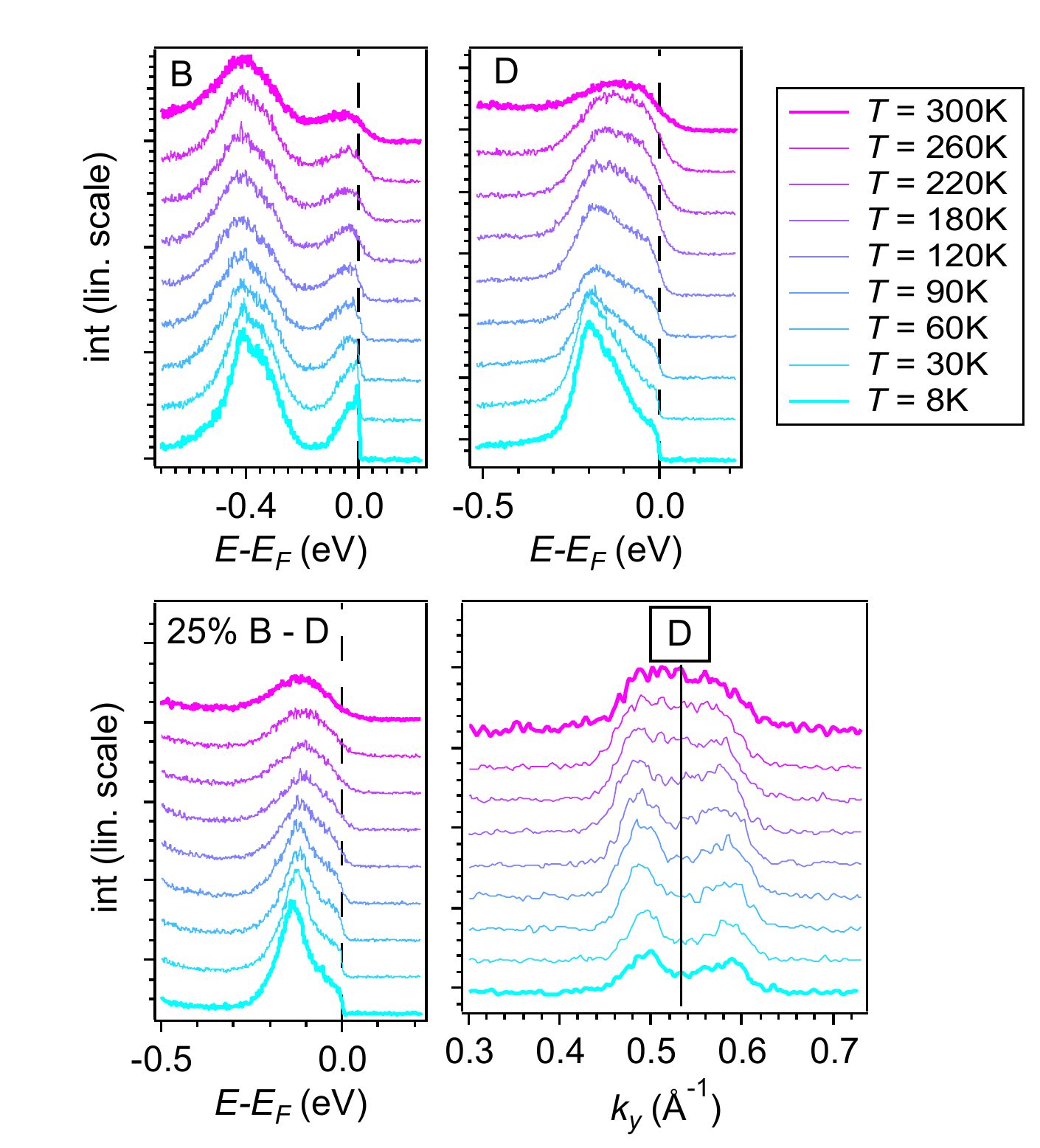}}
\caption{Temperature dependent ARPES spectra from \zrte\ at various positions in momentum space as indicated in each panel. The temperatures are listed in the legend. Data as a function of $E-E_F$ are energy distribution curves (EDC), which are taken precisely at B for the first panel and averaged over a small range of $k_y$ for the second and third panel. The fourth panel, as a function of $k_x$ shows momentum distribution curves (MDC) that are averaged over a 30~meV window around $E_F$. The spectra are vertically displaced for increasing temperature.}
\label{FigS2}
\end{figure*}

The CDW was previously associated with a temperature-dependent suppression of intensity of part of the 'q1D' Fermi surface sheet in angle-resolved photoemission spectroscopy (ARPES) experiments~\cite{yokoya05}. Here, we report data from \zrte\ aimed to reproduce the results of Ref.~[\onlinecite{yokoya05}] as well as data from \zrtese. The data from \zrte\ shown in Fig.~S\ref{FigS2} reproduce the main features of Ref.~[\onlinecite{yokoya05}], namely at B the emergence of a strong sharp peak at $E=E_F$ on lowering the temperature as well as the development of an intensity depletion at D in a range of approx $0<|E-E_F|<0.08$~eV. Even at the lowest temperature $T=8$~K a finite Fermi step at $E=E_F$ is still observed. Close to B at $k_x =0.2$~\AA$^{-1}$, which is 1/4 distance towards D a  sharpening of features but no intensity variation  at $E=E_F$ on lowering the temperature is observed. 

\begin{figure*}[h!]
\centerline{\includegraphics[height=14.8cm]{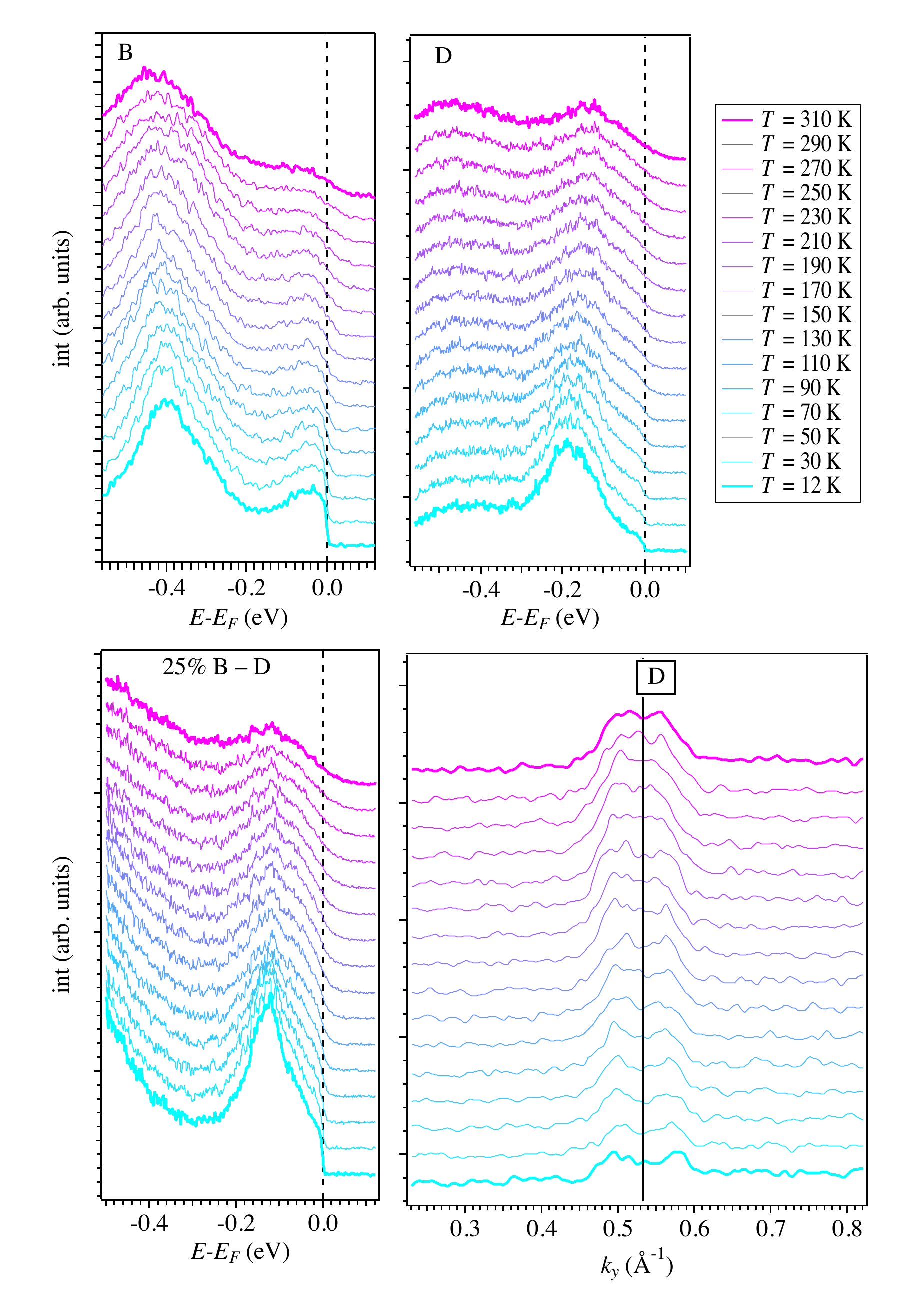}}
\caption{Temperature dependent ARPES spectra from \zrtese\ at various positions in momentum space as indicated in each panel. The temperatures are listed in the legend. Data as a function of $E-E_F$ are energy distribution curves (EDC), which are taken precisely at B for the first panel and averaged over a small range of $k_y$ for the second and third panel. The fourth panel, as a function of $k_x$ shows momentum distribution curves (MDC) that are averaged over a 30~meV window around $E_F$. The spectra are vertically displaced for increasing temperature.}
\label{FigS3}
\end{figure*}

The data from \zrtese\ in Fig.~S\ref{FigS3} show qualitatively similar behaviour to \zrte. The largest difference is observed at B, where the low energy peak is located at $E-E_F=0.03$~eV and no sharpening at low temperature is observed, thus leading to nearly constant intensity at $E=E_F$. A slight sharpening of the peak is still observed. At D an intensity depletion in the low energy region is observed, and a finite Fermi step height at the lowest temperature remains, similar to \zrte. The implications of this observation of spectral weight depletion are discussed in the main text. Near B, both for the band of the 3D Fermi surface sheet as well as for the q1D band regular behaviour without changes to the spectral weight at $E=E_F$ is found. 

\subsection{Discussion of Room Temperature Resistivity Anisotropy}

A remarkable feature of \zrte\ is the (accidental) in-plane near isotropy of resistivity $\rho_a/\rho_b\sim 1$ at room temperature (RT)~\cite{takahashi83}. This isotropy is lifted with an increase of $\rho_a$ and decrease of $\rho_b$~\cite{zhu16}. Our Fermi surface data, taken at low temperature, fail to rationalise this effect. In a simple Drude model, the resistivity can be estimated as 
\begin{equation}
\rho =\frac{m}{nq^2\tau},
\end{equation}
where $n$ is the carrier number, $m$ and $q$ are the electron mass and charge and $1/\tau$ is the scattering rate. Considering only the defect scattering, the observed increase of scattering rate as well as the decrease of FS volume in Se substituted samples would reduce the relative contribution of the 3D band to conductivity. The q1D band with its pronounced directionality contributes to conductivity along $a$ and less to conductivity along $b$, as manifest from the observation of a resistivity anomaly near $T_{CDW}$ in $\rho_a$ with no anomaly in $\rho_b$~\cite{takahashi83}. At RT, however, the phonon scattering as well as electron-electron scattering dominate over the defect scattering~\cite{jiang2011} and our low temperature results thus do not contribute any insight beyond the careful magneto-resistance analysis of Ref.~\cite{zhu16}.


\vspace{3mm}
\twocolumngrid

\bibliography{ArXiv_ZrTe3_disorder}

\end{document}